\documentclass{iopart}
\usepackage{graphicx}
\usepackage{revsymb}

\begin{document}
\title{A POSSIBILITY OF VOLUME REFRACTION OF NEGATIVE RELATIVISTIC PARTICLES IN BENT CRYSTALS}

\author{Gennady V. Kovalev}
\address{School of Mathematics \\University of Minnesota, Minneapolis, MN 55455,USA}

\begin{abstract}
The  volume coherent deflection of high-energy  positive and negative particles in uniformly bent crystals is studied. The general analysis of potential scattering shows that the standard screening potential for a large class of collisions can cause the volume refraction for negative particles (antiprotons, electrons) instead of the volume reflection for positive particles (proton, positrons).
\end{abstract}



\section{\label{sec:01}Introduction}
In the early 80-s, the experimental studies\cite{andreev_1982, andreev_1984, andreev_1986} have discovered the volume capture of relativistic particles in a channeling motion by bent crystals. Several attempts have been made to find a specific mechanism of capture(see e.g.\cite{tar98}) because the elastic scattering in a central field does not include the channeling trajectories (see below). The computer simulations of the passage of high energy particles in a uniformly bent crystal \cite{tar87_1,tar87_2} were completed and it was found that the particles (both positive and negative) are deflected to the side opposite the bend of atomic planes by an angle of about $2*\theta_c$ ($\theta_c$ - critical angle). It was called a ''volume reflection'' by bent atomic planes.  The significant beam divergence and multiple scattering of the quasi-channeled particles in the crystals did not allow to find this effect for almost two decades. The experiment \cite{ivanov_2005} \cite{ivanov_2006} of $70 GeV$ protons scattered by a short bent crystal is probably the first explicit demonstration of the volume reflection. Early, the preliminary results of crystal collimation at RHIC \cite{fliller_2005} for the 2001 run demonstrated some interested features which were later explained \cite{bir_2006_1, bir_2006_3} by the same effect. 

Our purpose is to discuss the principle difference between deflections of positive and negative particles if the standard continuous screening potential for the crystal planes  is considered to be a good approximation. In this approximation the sign of continuous screening potential is defined by the charge of projectile and the behavior of positive and negative particles are different. The positive particles should undergo the ''volume reflection'', but it  turns out that the  negative particles in such potential should deflect in the same direction as  the bend of atomic planes. It is not the channeling motion because the elastic scattering in central potential produces two symmetrical branches touching equipotential circle and extending from the turning point to infinity.  This effect is different from predicted in \cite{tar87_1,tar87_2} and can be called  a ''volume refraction'' by the bent atomic planes. 

\section{\label{sec:02}Screening Crystal Potential}
The charged particles incident on a single crystal with small angles to the crystallographic directions, experience the collective fields from atoms of slightly bent planes. The interaction potential for binary ion-atom collisions is, in general, described by 

\begin{eqnarray}
U_{atom}(r)=\frac{Z_P Z_T}{r}f(\frac{r}{R_{T-F}}),
\label{atom_potent}
\end{eqnarray}
where $Z_{P,T}$ are the bare nuclear charge of the projectile and crystal atoms, $R_{T-F}$ is  Thomas-Fermi (or Firsov) screening length and $f$ is a screening function ($f>0$). The screening function is 
\begin{eqnarray}
f(\frac{r}{R_{T-F}})=\sum_{i} \alpha_i exp(\beta_i\frac{ r}{R_{T-F}}),
\label{screening_factor}
\end{eqnarray}
and it approaches $1$ in the limit $r \rightarrow 0$ and decays 
exponentially for large $r$. This is only  possible if 
\[
\sum_{i} \alpha_i=1. 
\]
Different analytic forms of the screening function $f$ are in use. The 
Moliere approximation\cite{moliere_1947}, $f_M$, uses $\{\alpha_i\} = \{0.1,0.55,0.35\}$, 
$\{\beta_i\} = \{6.0,1.2,0.3\}$. Other choices include the Ziegler-Biersack-Littmark 
potential\cite{zbl_1985} $f_{ZBL}$ with $\{\alpha_i\} = \{0.1818,0.5099,0.2802,0.0281\}, \{\beta_i\} = \{3.2,0.9423,0.4028,0.2016\}$. Now, if we find the average potential over a cylinder with the radius $R >> R_{T-F}$ and the root-mean-square temperature  displacement $u$ of atoms in gaussian form, we receive the following expression for continuous cylindrical potential  
\begin{eqnarray}
U_{p}(\rho)=
\overline{U}\sum_{i}\frac{\alpha_i}{\beta_i}\exp(\beta_i^2\frac{u^2}{2R_{T-F}^2})
(\exp(\beta_i\frac{|\rho-R|}{R_{T-F}})erfc(\beta_i\frac{u}{\sqrt{2}R_{T-F}}+\frac{|\rho-R|}{\sqrt{2}u})\nonumber\\
+\exp(-\beta_i\frac{|\rho-R|}{R_{T-F}})erfc(\beta_i\frac{u}{\sqrt{2}R_{T-F}}-\frac{|\rho-R|}{\sqrt{2}u})).
\label{bent_PP}
\end{eqnarray}
Here $\rho$ is polar coordinate from center of cylinder, $\overline{U}=\pi Z_P Z_T R_{T-F} n_p$, $n_p$ is a density of atoms in the bent plane, $erfc()$ is the complementary error function\cite{abst_1964}. 
\begin{figure}[htbp]
	\centering
		\includegraphics{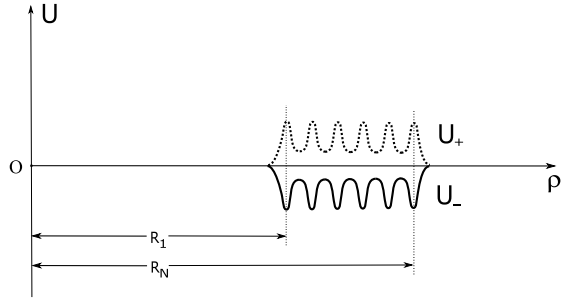}
	\caption{The radial profile of cylindrical potential.}
	\label{fig:PotetialCylinder}
\end{figure}

For positive ($Z_{P} > 0$) or negative ($Z_{P} < 0$) charged projectiles , the screen potential of atoms (\ref{atom_potent}) over all space ($0<r<\infty$) is repulsive ($U_{atom}(r)>0$) or attractive ($U_{atom}(r)< 0$) respectively. (Strictly speaking, the potential is called repulsive if the force is everywhere repulsive, i.e., the potential monotonely decreases to zero, but for simplicity we use the above terminology.) It follows that the average potential of each plane $U_{p}(\rho)$ and the whole cylindrical crystal potential

\begin{eqnarray}
U(\rho)=\sum U_{p}(\rho)
\label{bent_CP}
\end{eqnarray}
will also be repulsive or attractive depending on the charge of the projectile (Fig. \ref{fig:PotetialCylinder}). The summation in (\ref{bent_CP}) is over a range of radii $R_i, i=1..N$ which denote the centers of atomic planes located periodically from $R_{1}$ to $R_{N}$. For $0< \rho << R_{1}$ and $R_{N}<< \rho < \infty$ the potential decreases to zero. In the area $R_1 < \rho < R_{N} $, potential does not have a singularity and is smooth almost periodic function with some constant positive or negative pedestal. The value of pedestal is quite significant. It is  ~13.3 eV for Si $<110>$ plane (T=300K), i.e. about 1/3 of of maximal value of screening potential (~35.9 eV in this case). 

\section{\label{sec:03}Scattering in Central Field} 
\begin{figure}[htbp]
	\centering
		\includegraphics{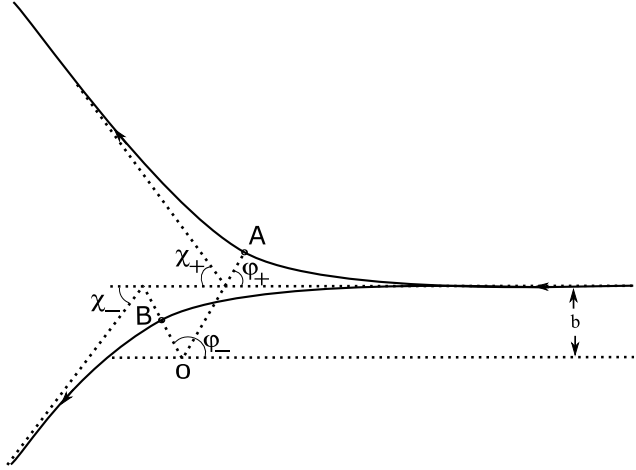}
	\caption{Scattering by repulsive and attractive potentials}
	\label{fig:ClassicScatter}
\end{figure}

The path of a particle in a central classical field is, as well known, symmetrical about a line from the center to the nearest point in the orbit (OA for positive and OB negative particles in Fig.\ref{fig:ClassicScatter}). 
Hence the two asymptotes to the orbit make equal angles ($\varphi_{+}$ for positive and $\varphi_{-}$ for negative particles) with these lines. The angles $\chi_{+,-}=\pi-2\varphi_{+,-}$ through which the positive and negative particles are deflected as they pass the center of potential have the different sign (see Fig.\ref{fig:ClassicScatter}), which we discuss below. This general rule is applied to any centrally (cylindrically) symmetrical potential including the cylindrical crystal with radius $R$. 

The solution for relativistic Hamilton-Jacobi equation in central field can be found  ( see e.g.\cite{newton_1982}) and it gives the well known expression for $\varphi$:     
\begin{eqnarray}
\varphi= \int^{\infty}_{\rho_{min}}\frac{\frac{b d\rho}{\rho^2}}{\sqrt{1-\frac{b^2}{\rho^2}-\frac{2U(\rho) E}{p_{\infty}^2}}},
\label{phi_angle}
\end{eqnarray}
where $E=\sqrt{p_{\infty}^2 c^2 + m^2c^4}$, $U(\rho)$ are the energy and potential of central field (\ref{bent_CP}); $b$, $\rho_{min}$, $p_{\infty}$ are the impact parameter, the nearest point in the orbit ($\rho_{min}$=OA or $\rho_{min}$=OB  in Fig.\ref{fig:ClassicScatter}) and the momentum of particle at infinity respectively.
If we use new variable $\varrho=\rho/b$, the scattering angle  (\ref{phi_angle}) can be written in more convenient form: 

\begin{eqnarray}
\varphi= \int^{\infty}_{\varrho_{min}}\frac{d\varrho}{\varrho\sqrt{(1-g(\varrho)) \varrho^2-1}},
\label{phi_angle2}
\end{eqnarray}
where $g(\varrho)=\frac{2U(b\varrho) E}{p_{\infty}^2 c^2}$. We should note that similarity of $g(\varrho)$ and the square of Lindhard angle $\theta_c^2=\frac{2\overline{U} E}{p_{\infty}^2 c^2}$ is not accidental.
The nearest point $\varrho_{min}$ is defined by the transcendental equation

\begin{eqnarray}
\varrho_{min}^2=\frac{1}{1-g(\varrho_{min})}.
\label{turning_point}
\end{eqnarray}
In the case of purely attractive potential, $g(\varrho)<0$, from (\ref{turning_point}) we can receive $\varrho_{min}<1$, i.e. the nearest point $\varrho_{min}$ is always less than impact parameter. In the case of purely repulsive potential $g(\varrho)>0$, we have opposite relation $\varrho_{min}>1$.
In the case of  $U(b\varrho)=0$, the nearest point $\varrho_{min}=1$ and the integral (\ref{phi_angle2}) becomes

\begin{eqnarray}
\varphi= \int^{\infty}_{1}\frac{d\varrho}{\varrho\sqrt{\varrho^2-1}}=\frac{\pi}{2},
\label{pi_angle}
\end{eqnarray}
which means that the particle does not deflect from straight line. 

\section{\label{sec:04}Classical Limits of Scattering Angle}
Now we can find the limits for classical variation of scattering angles $\varphi_{-},\varphi_{+}$. As it was pointed out for attractive potential ($g(\varrho)\leq0$), $\rho_{min}$ is less than impact parameter $b$ and the integral (\ref{phi_angle2}) can be estimated as

\begin{eqnarray}
\varphi_{-}= \int^{\infty}_{\varrho_{min-}}\frac{d\varrho}{\varrho\sqrt{(1-g(\varrho)) \varrho^2-1}} \geq \int^{\infty}_{\varrho_{min-}}\frac{d\varrho}{\varrho\sqrt{(1+g_{-}) \varrho^2-1}}=\nonumber\\
\arctan( \frac {1}{\sqrt {{\varrho_{min-}}^{2}+{\varrho_{min-}}^{2}g_{-}-1}} ).
\label{neg_case2}
\end{eqnarray}
Here $g_{-} =max(|g(\varrho)|), \varrho_{min-}\leq\varrho<\infty$ is a deepest value of the attractive potential (\ref{bent_CP}). The (\ref{neg_case2}) gives the lower bound of the scattering angle $\varphi_{-}$. For positive particles ($g(\varrho)\geq0$), it is easy to see that
\begin{eqnarray}
\varphi_{+}= \int^{\infty}_{\varrho_{min+}}\frac{d\varrho}{\varrho\sqrt{(1-g(\varrho)) \varrho^2-1}} \leq \int^{\infty}_{\varrho_{min+}}\frac{d\varrho}{\varrho\sqrt{(1-g_{+}) \varrho^2-1}}=\nonumber\\
\arctan( \frac {1}{\sqrt {{\varrho_{min+}}^{2}-{\varrho_{min+}}^{2}g_{+}-1}} ),
\label{pos_case}
\end{eqnarray}
where $g_{+} =max(g(\varrho)) \geq g(\varrho)$ is the highest point of repulsive potential. This is different from the scattering on repulsive  monotonely decreasing  potential \cite{newton_1982}, where this angle is always less than ${\pi}/{2}$.

\section{\label{sec:06}Small Angle Scattering}
Now let us turn to the small angle scattering in the central field (\ref{bent_CP}).
From physical pictures of scattering (Fig.\ref{fig:ClassicScatter}), it is clear that the particle's path has a biggest curvature near the turning point. So, the potential near that point plays a crucial role in the forming of scattering angles. In the rest of space the potential is analytic function, so we can introduce 'residual' potential $\overline{g}=g(\varrho_{min})-g$. Since $\overline{g}$ is small in the neighborhood of $\varrho_{min}$, we expand the square root $\sqrt{(1-g(\varrho_{min})+\overline{g})\varrho^2-1}$ (eq. (\ref{phi_angle2})) in powers of this residual potential:

\begin{eqnarray}
\varphi= \int^{\infty}_{\varrho_{min}}\frac{d\varrho}{\varrho\sqrt{(1-g(\varrho_{min}))\varrho^2-1}}-
\frac{1}{2} \int^{\infty}_{\varrho_{min}}\frac{ \overline{g}\varrho d\varrho}{((1-g(\varrho_{min}))\varrho^2-1)^{3/2}}\nonumber.
\label{phi_angle_small}
\end{eqnarray}

The first integral gives $\pi/2$. The second integral depends on the magnitude and  sign of scattering 'residual' potential,  as well as on the impact parameter and energy of particles. Its integrand has a divergent term in denominator at $\varrho=\varrho_{min}$. However, the 'residual' potential $\overline{g}(\varrho_{min})=0$, and there is no divergence in integral. The second integral can be integrated by parts, giving 

\begin{eqnarray}
\varphi= \frac{\pi}{2}-
\frac{1}{2} \int^{\infty}_{\varrho_{min}}\frac{d \overline{g}}{d \varrho} \frac{  d\varrho}{(1-g(\varrho_{min}))((1-g(\varrho_{min}))\varrho^2-1)^{1/2}}.
\label{phi_angle_small2}
\end{eqnarray}
Finally, using $\frac{d \overline{g}}{d \varrho}=-\frac{d g }{d \varrho}$ the formula for classical deflection angle for particles is

\begin{eqnarray}
\chi=\pi-2\varphi=
-\int^{\infty}_{\varrho_{min}}\frac{d g }{d \varrho} \frac{  d\varrho}{(1-g(\varrho_{min}))((1-g(\varrho_{min}))\varrho^2-1)^{1/2}}.
\label{final_angle_small}
\end{eqnarray}
The difference between deflection of positive and negative particles is seen from the following equations:
\begin{eqnarray}
\chi_{+}=
-\int^{\infty}_{\varrho_{min_{+}}}\frac{d g }{d \varrho} \frac{  d\varrho}{(1-g(\varrho_{min_{+}}))((1-g(\varrho_{min_{+}}))\varrho^2-1)^{1/2}},\label{final_angle_small+}\\
\chi_{-}=
+\int^{\infty}_{\varrho_{min_{-}}}\frac{d g }{d \varrho} \frac{  d\varrho}{(1+g(\varrho_{min_{-}}))((1+g(\varrho_{min_{-}}))\varrho^2-1)^{1/2}}.
\label{final_angle_small-}
\end{eqnarray}

\section{\label{sec:05}Conclusion}
The angle of scattering given by these equations is the first approximation in the expansion  (\ref{phi_angle_small}) in powers of the residual potential and can describe the major features of small angle deflection of positive and negative particles. It is more general than the classical small angle formula (see e.g. \cite{ll1}) in sense that it describes more accurate the turning points. They coincide if potentials have a long range of transversal variations and $\varrho_{min_{-,+}} \longrightarrow 1$, $g(\varrho_{min_{-,+}}) << 1$ in (\ref{final_angle_small+},\ref{final_angle_small-}). In this case we have
\begin{eqnarray}
\chi_{+}=
-\int^{\infty}_{1}\frac{d g }{d \varrho} \frac{  d\varrho}{(\varrho^2-1)^{1/2}},\\
\chi_{-}=
+\int^{\infty}_{1}\frac{d g }{d \varrho} \frac{ d\varrho}{(\varrho^2-1)^{1/2}}.
\label{final_angle_small3}
\end{eqnarray}

Since  $U_{+}(\rho)=-U_{-}(\rho)$, the deflective angles for positive and negative particles satisfy the following conditions:

\begin{eqnarray}
\chi_{+} =- \chi_{-}.
\label{angles}
\end{eqnarray}

So if the positive particles deflected by positive angle $\chi_{+}$, the negative particles should deflected to the opposite direction. For more accurate calculations one should use the formulas (\ref{final_angle_small+},\ref{final_angle_small-}).

\section{\label{sec:7}REFERENCES}

\bibliographystyle{unsrt}
\bibliography{../../Focusing_and_Channeling_in_Crystals/chan02}

\end{document}